\documentstyle[11pt,amssymb,epsf,cite]{article}
\input amssym.def
\input amssym.tex

\makeatletter
\@addtoreset{equation}{section}
\makeatother

\def\ben{\begin{equation}}
\def\een{\end{equation}}
\def\half{{1 \over 2}}
\def\bea{\begin{eqnarray}}
\def\eea{\end{eqnarray}}
\def \B{{\bf B}}

\def \p{\partial}

\def \A{{\bf A}}

\def\bnabla{{\bf \nabla}}

\def\ft#1#2{{\textstyle{{\scriptstyle #1}\over {\scriptstyle #2}}}}
\def\fft#1#2{{#1 \over #2}}

\def\half{{\textstyle{1\over2}}}
  \let\g=\gamma  
    
  \let\n=\nu \let\x=\xi \let\p=\pi 
 \let\t=\tau    
       \let\D=\Delta  
     
\let\C=\Chi

\def\nn{\nonumber} \def\bd{\begin{document}} \def\ed{\end{document}}
\def\ds{\documentstyle} \let\fr=\frac \let\bl=\bigl \let\br=\bigr
\let\Br=\Bigr \let\Bl=\Bigl 
\let\bm=\bibitem
\let\na=\nabla
\let\pa=\partial 
\let\ov=\overline 
\newcommand{\be}{\begin{equation}} 
\newcommand{\ee}{\end{equation}} 
\def\ba{\begin{array}}
\def\ea{\end{array}}
\def\ft#1#2{{\textstyle{{\scriptstyle #1}\over {\scriptstyle #2}}}}
\def\fft#1#2{{#1 \over #2}}
\def\del{\partial}
\def\vp{\varphi}
\def\sst#1{{\scriptscriptstyle #1}}
\def\oneone{\rlap 1\mkern4mu{\rm l}}
\def\td{\tilde}
\def\wtd{\widetilde}
\def\ie{\rm i.e.\ }
\def\dalemb#1#2{{\vbox{\hrule height .#2pt

        \hbox{\vrule width.#2pt height#1pt \kern#1pt
                \vrule width.#2pt}
        \hrule height.#2pt}}}
\def\square{\mathord{\dalemb{6.8}{7}\hbox{\hskip1pt}}}
\newcommand{\ho}[1]{$\, ^{#1}$}
\newcommand{\hoch}[1]{$\, ^{#1}$}  
\newcommand{\ra}{\rightarrow}
\newcommand{\lra}{\longrightarrow}
\newcommand{\Lra}{\Leftrightarrow}
\newcommand{\ap}{\alpha^\prime}
\newcommand{\bp}{\tilde \beta^\prime}
\newcommand{\tr}{{\rm tr} }
\newcommand{\Tr}{{\rm Tr} } 
\def\0{{\sst{(0)}}}
\def\1{{\sst{(1)}}}
\def\2{{\sst{(2)}}}
\def\3{{\sst{(3)}}}
\def\4{{\sst{(4)}}}
\def\5{{\sst{(5)}}}
\def\6{{\sst{(6)}}}
\def\7{{\sst{(7)}}}
\def\8{{\sst{(8)}}}
\def\n{{\sst{(n)}}}
\def\cA{{{\cal A}}}
\def\cF{{{\cal F}}}
\def\tV{\widetilde V}
\def\tW{\widetilde W}
\def\tH{\widetilde H}

\def\tE{\widetilde E}
\def\tF{\widetilde F}
\def\tA{\widetilde A}
\def\im{{{\rm i}}}
\def\jm{{{\rm j}}}
\def\km{{{\rm k}}}
\def\tY{{{\wtd Y}}}
\def\ep{{\epsilon}}
\def\vep{{\varepsilon}}
\def\R{\rlap{\rm I}\mkern3mu{\rm R}}
\def\bD{{{\bar D}}}
\def\C{{{\Bbb C}}}
\def\H{{{\Bbb H}}}
\def\RP{{{\Bbb R}{\Bbb P}}} 
\def\CP{{{\Bbb C}{\Bbb P}}} 
\def\HP{{{\Bbb H}{\Bbb P}}}
\def\Z{{\Bbb Z}} 
\def\bfe{{\bf e}}
\def\bfq{{\bf q}}
\def\cosec{\rm cosec}
\def\bB{\Bbb B}
\def\bA{\Bbb A}
\def\bC{\Bbb C}
\def\bZ{\Bbb Z}
\def\x{{\bf x}}
\def\g{\gamma}
\def\tx{\tilde x}
\def\tv{\tilde v}
\def\Dslash{\slash \negthinspace \negthinspace \negthinspace \negthinspace D}

\def\Aslash{\slash \negthinspace \negthinspace \negthinspace \negthinspace A}

\begin{document}

\def\D_Yslash{\slash \negthinspace \negthinspace 
\negthinspace \negthinspace D_Y}

\newcount\hour \newcount\minute
\hour=\time  \divide \hour by 60
\minute=\time
\loop \ifnum \minute > 59 \advance \minute by -60 \repeat
\def\nowtwelve{\ifnum \hour<13 \number\hour:
                      \ifnum \minute<10 0\fi
                      \number\minute
                      \ifnum \hour<12 \ A.M.\else \ P.M.\fi
	 \else \advance \hour by -12 \number\hour:
                      \ifnum \minute<10 0\fi
                      \number\minute \ P.M.\fi}
\def\nowtwentyfour{\ifnum \hour<10 0\fi
		\number\hour:
         	\ifnum \minute<10 0\fi
         	\number\minute}
\def\now{\nowtwelve}

\voffset=-0.8in

\hfuzz=100pt

\title{Time-Dependent Multi-Centre Solutions 
from New Metrics with  Holonomy ${\rm Sim}(n-2)$}
\author{ \Large G.W. Gibbons$^{(1)}$ and C.N. Pope$^{(2,1)}$
\\
\\
\\
$^{(1)}$D.A.M.T.P.,
 Cambridge University,
\\ Wilberforce Road,
 Cambridge CB3 0WA,
U.K.
\\
\\ 
\\ $^{(2)}$ George P. \& Cynthia W. Mitchell Institute for Fundamental Physics,
\\Texas A\&M University,
College Station,
\\  TX 77843-4242,
USA
\\
\\
\\
}

\maketitle

\vskip- 3.1cm
\medskip

	 \newdimen\crestht
	 \crestht=60 pt

	 \epsfysize=\crestht
{\rightline{\epsfbox{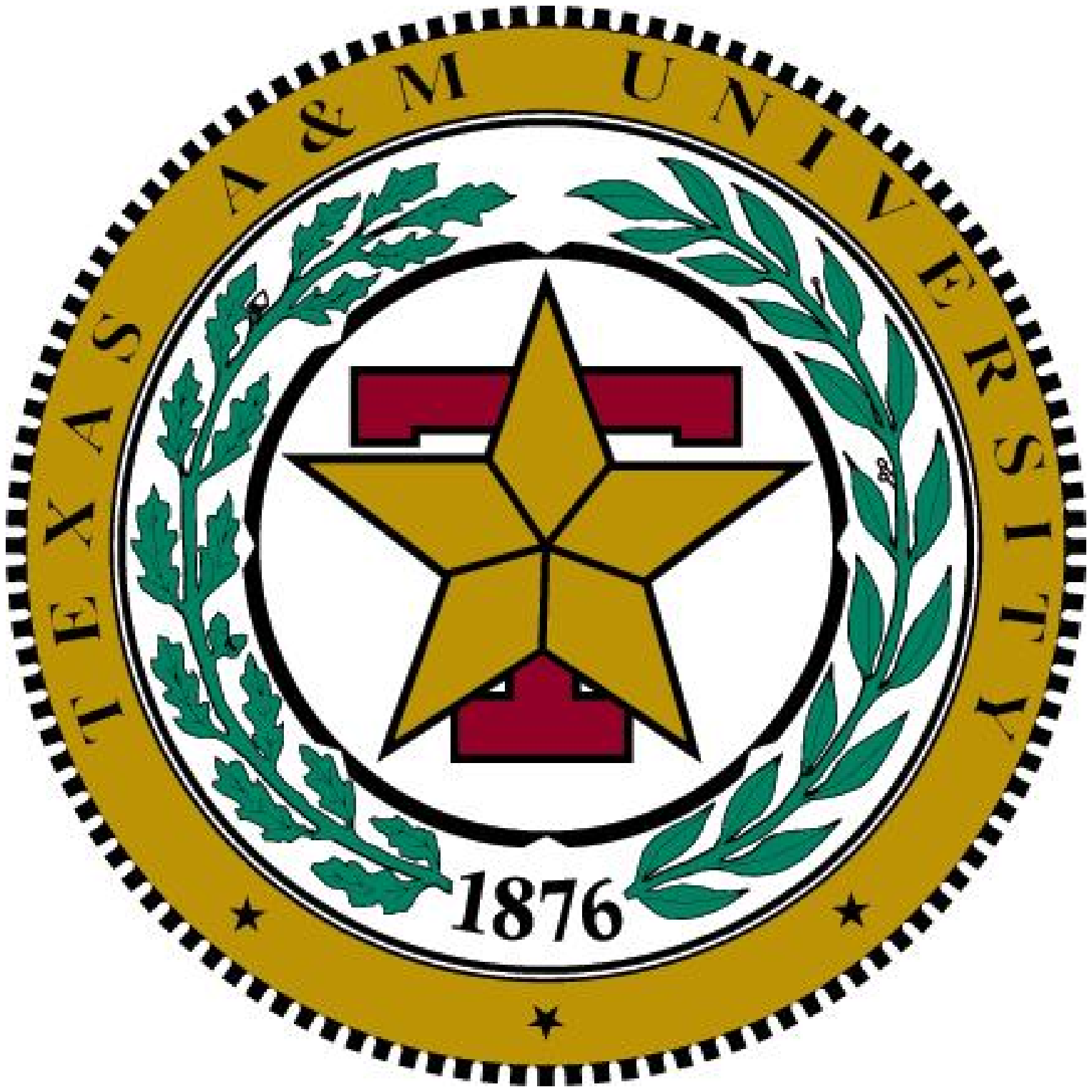 }
}}
\vskip -2.0cm\epsfysize=\crestht
{\leftline{\epsfbox{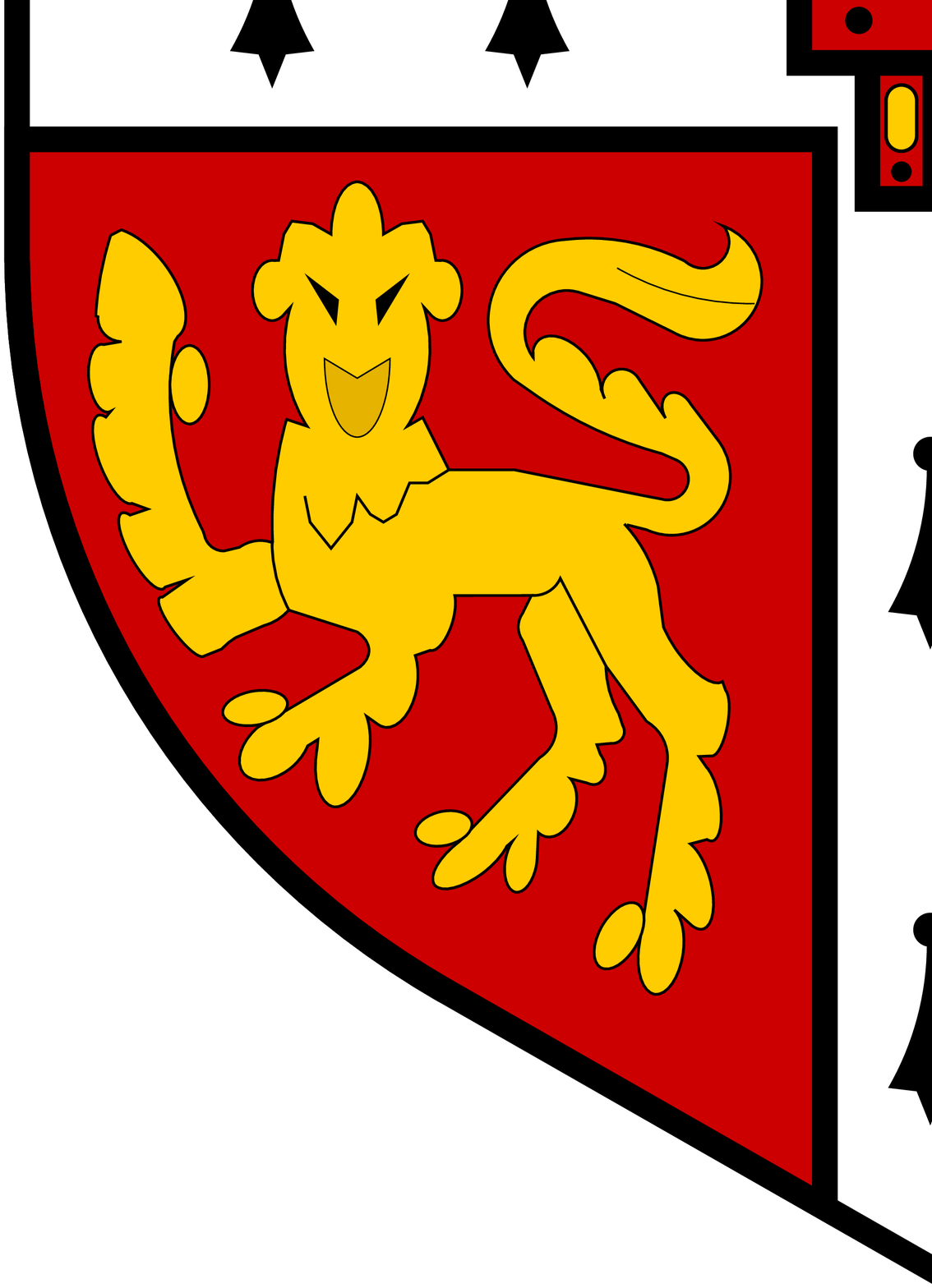}
}}

\begin{abstract} 

   The classifications of holonomy groups in Lorentzian
and in Euclidean signature are quite different.  A group of
interest in Lorentzian signature in $n$ dimensions is the maximal
proper subgroup of the Lorentz group, $\rm Sim(n-2)$. 
 Ricci-flat
metrics with $\rm Sim(2)$ holonomy were  constructed by Kerr and Goldberg, 
and a single
four-dimensional example with a non-zero cosmological constant was exhibited
by Ghanam and Thompson. Here  we reduce  the problem of finding 
the general $n$-dimensional Einstein metric of 
$\rm Sim(n-2)$ holonomy, with
and without a cosmological constant, to solving a set linear 
generalised Laplace and Poisson equations on an $(n-2)$-dimensional
Einstein base manifold.  Explicit examples may be constructed
in terms of generalised harmonic functions. 
A dimensional reduction of these multi-centre solutions 
gives  new 
time-dependent Kaluza-Klein black holes and monopoles, including time-dependent
black holes in a cosmological background whose  spatial sections
have  non-vanishing curvature.

\end{abstract}

\vskip0.5in
\leftline{DAMTP-2007-88 \ \ \ \  MIFP-07-24}

\thispagestyle{empty}

\pagebreak
\voffset=0pt

\setcounter{page}{1}

\tableofcontents

\addtocontents{toc}{\protect\setcounter{tocdepth}{2}}

\newpage

\section{Introduction}

Apart from their obvious intrinsic interest,
solutions of Einstein's equations, vacuum or with 
supergravity sources, representing 
arbitrary many black  holes, naked singularities
or p-branes  moving 
in a time-independent or 
time-dependent background have found many applications 
in string and M-theory. Such metrics are often called 
multi-centre metrics because they depend on one or more arbitrary
functions which are harmonic with  respect to some transverse spatial metric 
$g_{ij}$. In this sense the Einstein equations linearise and 
one may therefore  take  a superposition of solutions
with delta-function sources. 

A detailed examination of these `anti-gravitating' \cite{Scherk} 
solutions
near the sources often reveals that they can be regarded
as extreme  black holes \cite{HartleHawking}, or products of anti-de-Sitter
spacetime with an Einstein space \cite{GibbonsTownsend},
or as singular extreme limits of  black holes \cite{GibbonsTownsend}.
In other cases the delta-functions represent magnetic monopoles
\cite{Hawking,GibbonsHawking,GrossPerry,Sorkin} and may be resolved
by passing to higher dimensions \cite{GibbonsHorowitzTownsend}.
In some  cases the presence of more than one centre
gives rise to new singularities \cite{GibbonsHorowitzTownsend,
Welch,CandlishReall}. 
The situation  most studied is when the transverse metric $g_{ij}$
is time independent and flat. The harmonic functions
may then be taken to be a superposition of solutions of
Green functions for the time-independent Laplace equation on
Euclidean  space. In many cases these metrics may be generalised
to the case where the transverse space is curved but still Ricci flat. 
However, so
far we know of few  solutions of this type for which
$g_{ij}$ is an Einstein space
\ben    
R_{ij}= \Lambda g_{ij}\,,
\een
The main example is the Anti-De-Sitter
generalisation of the Brinkmann solutions given  in \cite{GibbonsRuback}. 
These are of the form \footnote{The full n-dimensional metric
satisfies $R_{\mu \nu}=\Lambda g_{\mu \nu}$.} 
\ben
ds ^2 = {n-1 \over (-\Lambda y^2)} 
\Bigl \{ 2 dudv + H(u,y,x^a)  du ^2 + dy ^2 + h_{ab} dx^a dx ^b \Bigr \}
\een
with $a=1,2,\dots,(n-3)$, the metric $h_{ab}(x)$ being Ricci flat
and hence the transverse metric
\ben
g_{ij}dx^i dx ^j= {n-1 \over (-\Lambda  y^2)} 
\Bigl \{ dy^2 + h_{ab} dx^a dx^b \Bigr \}  
\een
being an Einstein metric with negative scalar curvature.

The function  $H$ satisfies the  Laplace type equation 
\ben
y^{n-2} \pa_ y \bigl ({1 \over y^{n-2}}  \pa _y H  \bigr) + 
\nabla ^2_h H =0\,, 
\een
$\nabla ^2_h$ being the Laplacian with respect to the
metric $h_{ab}$.

Another  example  is  the  multi-domain wall solution \cite{CsakiShirman}
which, from the point of view of this paper, is
rather degenerate case, since it may more simply  be viewed as 
patches of $AdS$ glued together  across hypersurfaces.

As well as time-independent solutions, a number
of time-dependent solutions are known 
(see 
\cite{KastorTraschen,MakiShiraishi,GibbonsLuPope,ChenChongGibbonsLuPope}),
but as far as we know
so far these are all in the case
where the transverse  metric is time independent and Ricci flat.

   Time-independent multi-centre metrics may (or may not)
be supersymmetric or BPS. That is, considered
as solutions of a supergravity theory they may admit Killing spinors
\cite{GibbonsMunich,GibbonsTrieste,GibbonsLondon,GibbonsHull}.
Thus, in the vacuum case, they may admit a covariantly-constant commuting 
spinor field $\epsilon$.  
Time-dependent solutions cannot be supersymmetric,
since any Killing spinor field $\epsilon$ gives rise
to a non-spacelike Killing vector field $\bar \epsilon \Gamma ^\mu \epsilon$.
However, it can happen \cite{Freedman,GibbonsLuPope,ChenChongGibbonsLuPope} 
that  the time-dependent solution
arises from the dimensional  reduction of a time-independent 
solution in one higher
dimension that does admit a Killing spinor, and hence
a non-spacelike Killing vector field   
$\bar \epsilon \Gamma ^\mu \epsilon $ and that also admits an 
 additional, spacelike,
boost Killing vector field.
Dimensional reduction with respect to this boost Killing field
then gives rise to a time-dependent solution
in the lower dimension. Because the Killing spinor $\epsilon$
is not boost invariant, it does not descend to the
lower-dimensional spacetime, which is therefore, unlike its higher-dimensional
progenitor, not supersymmetric.

Multi-centre metrics can  have, or arise by dimensional reduction from,
a metric $g_{\mu \nu}$, $\mu,\nu =1,2,\dots ,n$ 
with a reduced holonomy group. Thus for example
in the case that the Killing spinor is covariantly constant
$\nabla _\mu \epsilon =0$, the associated Killing vector
$n^\mu = \bar \epsilon \Gamma ^\mu \epsilon$
is also covariantly constant, $\nabla_\mu  n^\nu =0$. 
A particularly interesting example is when the Killing vector
field is null $n^\mu n_\mu =0$.  The subgroup of the Lorentz group
$SO(n-1,1)$ leaving the  null vector $n^\mu$ invariant
is the   Euclidean group $E(n-2)$. Thus a metrici admitting
a covariantly constant Killing vector has  holonomy group
$E(n-1)$ or a subgroup thereof. 
Such metrics are known as Brinkmann waves 
\cite{Brinkmann1,Brinkmann2,Brinkmann3}.
A special case are the so-called pp-waves of the form
\ben
ds ^2 = 2 dudv + dx^i dx ^i + H(u,x^i) du^2\,,
\een
where the Ricci flat condition becomes
\ben
\partial_i \partial _i H =0\,.
\een
Such metrics   represent
gravitational radiation propagating in one fixed 
direction, and in fact have holonomy consisting of just the
translation subgroup ${\Bbb R} ^{n-2}$ of $E(n-2)$.

A spacetime admitting a covariantly constant null vector field is  also said to 
admit a Bargmann structure because
dimensional reduction on a covariantly-constant  null Killing vector field
$n^\mu$  gives rise to a  non-relativistic 
Newton-Cartan spacetime with  
a degenerate co-metric \cite{Duval,DuvalGibbonsHorvathy}. 
Dimensional reduction on a null Killing vector that is
not necessarily covariantly constant has been studied by \cite{JuliaNicolai}.

  Brinkmann waves necessarily have vanishing Ricci scalar $R^\mu_\mu =0$.
They can be used to obtain time-independent \cite{GibbonsNP}
and time-dependent \cite{GibbonsLuPope}  
extremal Kaluza-Klein multi black holes, 
and by electric-magnetic duality, multi Kaluza-Klein monopole
metrics with Euclidean transverse space sections.
However, these extremal Kaluza-Klein black holes or multi Kaluza-Klein 
monopoles move in a cosmological 
Friedman-Lemaitre-Robertson-Walker background with 
spatial  sections having a flat metric $g_{ij}$. 
Interestingly the $AdS_n$ analogue of pp-waves, while
admitting a Killing spinor, and hence being supersymmetric
in the appropriate dimensions, have the full $SO(n-1,1)$ as
holonomy group. 

If we are to obtain solutions with  non-vanishing spatial curvature
from Einstein metrics in one higher dimension with reduced
holonomy and 
with non-vanishing cosmological constant,
we need to look at a   holonomy group which is larger than the
that  of the Brinkmann waves.
Now, the maximal proper subgroup of the Lorentz group
$SO(n-1,1)$ is the $1+\half(n-2)(n-3)$ subgroup
${\rm Sim} (n-2)$, which leaves invariant a null 
direction\footnote{Some more details about ${\rm Sim}(n-2)$ are given in 
the appendix.}.

Thus we need to find examples of Einstein metrics $g_{\mu\nu}$
such that $R_{\mu \nu} =\Lambda g_{\mu \nu}$ and  with
holonomy ${\rm Sim} (n-2)$, and especially, those with
non-vanishing cosmological constant $\Lambda$.
This problem appears not to have been greatly studied.

In four dimensions, $n=4$, and if the Einstein equations
hold, holonomy ${\rm Sim}(2)$ implies that the Weyl tensor
is of Petrov-Plebanski type $III$,  whilst holonomy
${\Bbb R}^2 $ implies that it is of type $N$
\cite{Schell}.  The latter can occur only if $\Lambda = 0$.
Holonomy ${\rm Sim} (2)$ can occur both for $\Lambda=0$ and for
$\Lambda \ne 0$; an example of the latter is presented in
\cite{Ghanam}\footnote{Note, however, that the metric presented in
\cite{Ghanam} has misprints; see later.}. All solutions with
 $\Lambda=0$ and holonomy ${\rm Sim} (2)$ have been found, up to
a solution of two linear equations \cite{GoldbergKerr1,GoldbergKerr2}.
In the  literature \cite{Schell}, the Lie algebras
$\frak{sim}(2)$ and
${\Bbb R} ^2$, thought of as sub-algebras of the Lorentz algebra
$\frak{so}(3,1)$, are sometimes denoted by $R_{14}$ and $R_7$ respectively.
In five dimensions, all vacuum metrics with
$E(3)$  holonomy, or a subsgroup thereof,
 were written down by Brinkmann \cite{Brinkmann3}.
They depend on two functions that are harmonic in the
three transverse variables but which are otherwise arbitrary.

     If the functions in the five-dimensional Brinkmann metrics 
are taken to depend only on the transverse variables, 
one may dimensionally reduce to four dimensions
and obtain stationary solutions of  dilaton Einstein Maxwell gravity with
dilaton-Maxwell coupling characterised by $a=\sqrt3$ \cite{GibbonsNP}.
The case with for which   $A_i=$ was used in \cite{GibbonsNP} to
construct multi-electrically charged  extreme
black-hole solutions. The electric-magnetic dual
then yielded the {\sl Riemannian}  multi-Taub-NUT solutions
\cite{Hawking,GibbonsHawking} which may be interpreted as multi 
Kaluza-Klein monopoles \cite{GrossPerry,Sorkin}.

  The case with $A_i \ne 0$, which requires two harmonic functions,
was reduced to four dimensions to obtain multi- {\sl Lorentzian}
Taub-Nut solutions \cite{DuvalGibbonsHorvathy}.

   Another type of reduction was introduced in
\cite{Freedman}, in which the $u$-dependence is non-trivial
but chosen so as to make the metric invariant under the
the $SO(1,1)$ action
$u \rightarrow \lambda u$, $v \rightarrow \lambda^{-1}  v$.
Reduction on the boost Killing vector gives {\sl time-dependent}
multi-centre metrics  of a type first constructed by Kastor and Traschen
\cite{KastorTraschen} and generalised by Maki and Shiraishi
\cite{MakiShiraishi}. These have recently been used to
discuss the collisions of branes \cite{GibbonsLuPope,ChenChongGibbonsLuPope}.
In particular, five-dimensional Brinkmann waves
were used in \cite{GibbonsLuPope} to construct
time-dependent multi Kaluza-Klein monopole solutions.

    In this paper we shall reduce the general problem
of finding  $n$-dimensional Einstein metrics with holonomy ${\rm Sim} (n-2)$ 
to solving a linear system of Laplace-like and Poisson-like linear
equations on an $(n-2)$-dimensional transverse 
Einstein metric $g_{ij}( x^k, u)\,,\quad R_{ij}=\Lambda g_{ij}$, which can in
general be time dependent.
We give some explicit  examples and discuss their dimensional 
reductions to give time-dependent extremal Kaluza-Klein black holes
moving in a Friedman-Lemaitre-Robertson-Walker  
background with curved spatial sections and dominated by
a scalar field with a Liouville potential.    

   It is striking that despite the historical sequence 
in which examples in four dimensions
 with ${\rm Sim}(2)$  holonomy were found, it is actually 
quite a lot simpler to obtain examples with $\Lambda$ non-zero than it 
is to obtain $\Lambda=0$  examples.

The paper is organised as follows. In section 2 we describe the relationship
between ${\rm Sim}(n-2)$ holonomy and the existence  of a recurrent null
vector, leading us to a local form of the most general   
metric with this holonomy which was first written down by A G Walker. 
In section 3 we discuss the implications
of   ${\rm Sim}(n-2)$ holonomy for the existence of special
kinds of spinor fields. In section 4 we impose the Einstein equations
on the Walker metrics 
and reduce them to solving a linear system of equations in
in the $(n-2)$-dimensional transverse metric with its Einstein metric.
In section 5, after describing the general reduction technique,
 we use  the solutions we have obtained to 
construct time-dependent multi-centre metrics in 3+1 spacetime
dimensions. In section 6 we discuss the circumstances
under which the  time dependence of the transverse metric
may be eliminated. Our conclusions are contained in section 7.
The appendix contains a short, self-contained and  unified description of
the Galilei, Bargamnn, Carroll and Sim and ISim groups
appearing in the paper  
as subgroups of a higher dimensional Poincare group.

\section{Holonomy ${\rm  Sim}(n-2)$ and recurrent null vector fields}

   A metric with 
holonomy ${\rm Sim}(n-2)$ is by definition one which 
admits  a null vector field
$n^\mu$, $g_{\mu \nu} n^\mu n^\nu=0$,  
whose direction remains invariant under parallel transport.
This means that the null vector $n^\mu$ is `recurrent,' i.e.
it satisfies  
\ben
\nabla _\mu n^\nu = B_\mu n^\nu\,, \label{recurrent}
\een
for some `recurrence 1-form' $B_\mu$.
Note that there is a gauge freedom, since
$n^\mu$  and $\Omega n^\mu$ define the same  
null direction field. Under such a rescaling the recurrence form
changes as 
\ben
B \rightarrow B - d \Omega\,.
\een
 
   If we use the metric to convert $n^\mu$ to a 1-form
$n_\mu dx^\mu = g_{\mu \nu} n^\mu d x^\nu$, we may,   
by skew-symmetrising (\ref{recurrent} ), deduce that 
\ben
dn=B\wedge n\,, \label{ext}
\een
and hence that
\ben
n \wedge dn =0\,.
\een
It follows from Frobenius' theorem (see, for example, \cite{exact}) that
\ben
n= f du\,,
\een
for functions $f$ and $u$. Using the rescaling freedom
we may set $f=1$ and thus
\ben
 n=du \,, \qquad dn=0\,.
\een
Now (\ref{ext})  implies that $B= \kappa n$ for some function $\kappa$, 
and so (\ref{recurrent}) becomes 
\ben
\nabla _\mu n_\nu = \kappa n_\mu n_\nu \,.
\een
It follows that $n^\mu$ is tangent to
an affinely-parameterised, non-expanding, twist-free   
null geodesic congruence\footnote{This distinguishes
metrics of ${\rm Sim}(n-2)$ from the much less rich 
 higher-dimensional Robinson-Trautman solutions,
which admit an expanding twist-free shear-free null geodesic 
congruence \cite{Ortaggio}.}. The null vector field
$n^\mu$ is tangent to the null generators of the
null hypersurfaces $ u={\rm constant}$.

We may now introduce as coordinates the function $u$ and the affine 
parameter $v$, such that
\ben
n^\mu {\partial \over \partial x^\mu} = {\partial \over \partial v}\,,
\een
together with 
$n-2$ transverse coordinates $x^i$, $i=1,2,\dots,n-2$.  
The metric may now be cast into the form introduced by 
Walker\footnote{Not to be
confused with the Friedman-Lemaitre-Robertson-Walker metrics.} \cite{Walker}: 
\ben
ds ^2 = 2 du dv +g_{ij} (u,x^k) dx ^i dx ^j + H(u,v,x^i) du^2
+ 2 A_i (u, x^j)  dx^i du\,.\label{walkermet}
\een 
where $g_{ij}(u,x^k)$, $A_i(u, x^k)$, and $H(u,v,x^i)$ are 
arbitrary functions of their arguments.   We shall  
shortly constrain these functions by imposing the Einstein vacuum 
field equations
with cosmological term, $R_{\mu \nu}= \Lambda g_{\mu \nu}$.

     Defining $x^+ = v$ and $x^-=u$,  implying $n^\mu = \delta ^\mu _+$ and
$n_\mu =  \delta _\mu ^-$, and denoting partial derivatives by
\be
f'\equiv \fft{\del f}{\del v}\,,\quad \dot f \equiv
 \fft{\del f}{\del u}\,,\quad
\del_i f \equiv \fft{\del f}{\del x^i}\,,
\ee
one finds that
\ben
\kappa = \half H' \,,
\een 
and so
\be
\nabla_\mu n_\nu = \ft12 H'\, n_\mu n_\nu\,.\label{nabn}
\ee
The  Ricci identity 
\ben
\bigl ( \nabla_ \mu \nabla _\nu -\nabla _\nu \nabla _\mu \bigr) 
 n_\tau = R_{\tau \sigma  \mu \nu} n^\sigma
\een
then implies  that 
\ben
R_{i+\mu \nu}= 0\,\qquad R_{-+-+} = \half H''\,.
\een

    The special case where $H$ does not depend on the affine parameter $v$,
and so $\kappa =\ft12 H'=0$, corresponds to Brinkmann  waves
\cite{Brinkmann1,Brinkmann2,Brinkmann3}. In this case
the null vector field $n^\mu$ is covariantly constant and 
hence $ {\partial \over \partial v}$ becomes a null Killing vector
field. The holonomy group is then  reduced further, 
to $E(n-2)$ or a subgroup thereof.   

\section{Holonomy and Spinors}

\subsection{Calculation of local holonomy}

   The local holonomy algebra is generated by the curvature tensor of the
metric.  The algebra is most easily seen if one uses a vielbein basis for
the Riemann tensor.  For the Walker metrics (\ref{walkermet}) a suitable
basis, in which the metric takes the form $ds^2 = 2 \hat e^+ \hat e^-
  +\hat e^\alpha \hat e^\alpha$, is given by
\bea
\hat e^+ &=& dv + A_i dx^i + \ft12 H du\,,\nn\\
\hat e^- &=& du\,,\nn\\
\hat e^\alpha &=& e^\alpha\,,\label{vielbein}
\eea
where $e^\alpha$ is a vielbein for the transverse metric, $e^\alpha e^\alpha
= g_{ij} dx^i dx^j$.  It is useful also to record that the inverse vielbein
is given by
\bea
\hat E_+ &=& \fft{\del}{\del v}\,,\nn\\
\hat E_- &=& \fft{\del}{\del u} - \ft12 H\, \fft{\del}{\del v}\,,\nn\\
\hat E_\alpha &=& E_\alpha - A_\alpha\, \fft{\del}{\del v}\,.
\eea

   We shall just present the expressions that give the non-vanishing
vielbein components $\hat R_{abcd}$ of Riemann tensor of (\ref{walkermet})
in the case where the functions
$H$, $g_{ij}$ and $A_i$ in (\ref{walkermet}) are taken to be independent
of $u$.  We then find
\bea
\hat R_{+-+-} &=& -\ft12 H''\,,\nn\\
\hat R_{-\alpha + -} &=& \ft12(\nabla_\alpha H' - A_\alpha H'')\,,\nn\\
\hat R_{-\alpha-\beta}&=& -\ft12 \nabla_\alpha\nabla_\beta H -
      \ft12 H'' A_\alpha A_\beta + \ft14 F_{\alpha\gamma} F_\beta{}^\gamma 
+ \ft14 H'\, (\nabla_\alpha A_\beta + \nabla_\beta A_\alpha) \nn\\
&&+
   \ft12(A_\alpha \nabla_\beta H' + A_\beta\nabla_\alpha H')\,,\nn\\
\hat R_{-\alpha\beta\gamma} &=& \ft12 \nabla_\alpha F_{\beta\gamma}\,,\nn\\
\hat R_{\alpha\beta\gamma\delta} &=& R_{\alpha\beta\gamma\delta}\,.
\label{Riemann}
\eea
Note, in particular, that all components of the form $\hat R_{+\alpha  bc}$
vanish, where $b=(+,-,\beta)$ and $c=(+,-,\gamma)$.
This means that the local holonomy
is generated just by the $\rm Sim(n-2)$ subset of the Lorentz generators
$M_{ab}$, comprising
\be
M_{+\alpha}\,,\qquad M_{+-}\,,\qquad M_{\alpha\beta}\,.
\ee
(In other words, all generators except $M_{-\alpha}$.)

   It is clear that in the more complicated case where $H$, $A_i$ and
$g_{ij}$ in (\ref{walkermet}) are allowed also to depend on $u$ the
set of non-vanishing Riemann tensor components will be at least as
large as in the $u$-independent case (\ref{Riemann}) that we have
presented explicitly.  On the other hand, it is also clear that the components
$\hat R_{+\alpha  bc}$ will continue to vanish when $u$-dependence is
included.  This follows from recurrence condition $\nabla_\alpha n_\beta
= \ft12 H'\, n_\alpha n_\beta$, where the null vector $n$ is given by
$n=du= e^-$, by applying a second derivative and forming
a commutator, as discussed previously.  Therefore the local holonomy is
again $\rm Sim(n-2)$ for the general Walker metrics (\ref{walkermet})
that include $u$-dependence.

\subsection{Spinors in $\rm Sim(n-2)$ metrics}

     In any of the metrics (\ref{walkermet}) of $\rm Sim(n-2)$ holonomy
there exists a preferred spinor which generalises the covariantly-constant
spinor that exists in the Brinkmann wave metrics. In the Brinkmann case,
if the Einstein vacuum equations with $\Lambda=0$ hold within the
framework of a supergravity theory, then this spinor
generates half of the maximum number of supersymmetries. By contrast,
although there still exists a preferred spinor in the more general
$\rm Sim(n-2)$ holonomy metrics we are considering, this is not
associated with any supersymmetry.

   In the vielbein basis
(\ref{vielbein}), a calculation of the spin connection shows that the
Lorentz-covariant exterior derivative acting on spinors is given by
\bea
\hat D &\equiv & d +\ft14 \hat\omega_{ab} \Gamma^{ab} \nn\\
&=& d + \ft14 \omega_{\alpha\beta}\Gamma^{\alpha\beta}
+(\ft14 F_{\alpha\beta} -\ft14 \dot g_{ij} E^i_\alpha E^j_\beta)
      \Gamma^{-\alpha} \, \hat e^\beta\\
&&
\!\!\!\!\!\!\!+
\Big[(\ft14 \nabla_\alpha H -
  \ft14 H' A_\alpha -\ft12 \dot A_i E^i_\alpha)\Gamma^{-\alpha} -
\ft14 H' \Gamma^{+-}+
  \ft18(B_{\alpha\beta}-F_{\alpha\beta})\Gamma^{\alpha\beta}\Big]\hat e^-
\,,\nn
\eea
where
\be
B_{\alpha\beta} \equiv \ft12(\dot e_{i\beta} E^i_\alpha - \dot e_{i\alpha}
   E^i_\beta)\,.
\ee

   It is now evident that if we consider a spinor $\ep$ that satisfies
\be
\Gamma^-\ep=0\,,
\ee
and that is independent of $u$ and $v$, then the Dirac operator $\Gamma^a
\hat D_a$ acting on $\ep$ reduces to the Dirac operator
$\Gamma^\alpha D_\alpha$ in the internal space:
\be
\Gamma^a \hat D_a\ep = \Gamma^\alpha D_\alpha \ep\,,
\ee
where $D\equiv d + \ft14 \omega_{\alpha\beta}\Gamma^{\alpha\beta}$.  In
particular, if the internal space is maximally-symmetric, with
\be
R_{\alpha\beta\gamma\delta}= \fft{\Lambda}{n-3}\, (\delta_{\alpha\gamma}
\delta_{\beta\delta}-\delta_{\alpha\delta} \delta_{\beta\gamma})\,,
\ee
then it admits a Killing spinor $\ep$ satisfying
\be
D_\alpha\ep = \sqrt{-\fft{\Lambda}{4(n-3)}}\, \Gamma_\alpha\ep\,.
\ee
This spinor therefore satisfies a massive Dirac equation in the full
$n$-dimensional metric of $\rm Sim(n-2)$ holonomy, with
\be
\Gamma^a \hat D_a\ep = (n-2)  \sqrt{-\fft{\Lambda}{4(n-3)}}\, \ep\,.
\ee

  The spinor $\ep$, which we take to be commuting,
can be viewed as the square root of the distinguished
null vector $n^\mu$, and with a suitable normalisation we have
\be
n^\mu = \bar\ep\Gamma^\mu \ep\,.
\ee

\section{Einstein Equations and New $\rm Sim(n-2)$ Holonomy Metrics}

\subsection{The Einstein equations}

   After some algebra, we find that the non-vanishing 
coordinate-frame components $\hat R_{\mu\nu}$ of the
Ricci tensor of
the general Walker class (\ref{walkermet}) of $\rm Sim(n-2)$
holonomy metrics are given by
\bea
\hat R_{--} &=& -\ft12\Big[ \nabla^2 H + \del_u(g^{ij} \dot g_{ij}) -
   \ft12 F^{ij}F_{ij} +\ft12 \dot g^{ij} \dot g_{ij} + g^{ij}\ddot g_{ij}
-2\nabla^i \dot A_i\nn\\
&& - 2 A^i \del_i H' -(\nabla^i A_i)\, H' -H H'' +
\ft12 g^{ij} \dot g_{ij} H' + A^i A_i H''\Big]\,,\\
\hat 
R_{-i}&=& \ft12 \del_i H' +\ft12 \nabla^j F_{ij} + \ft12 \nabla^j \dot g_{ij}
      -\ft12 \del_i(g^{jk} \dot g_{jk})\,,\\
\hat R_{+-} &=& \ft12 H''\,,\\
\hat R_{ij} &=& R_{ij}\,,\label{riccigen}
\eea
where we have defined
\be
F_{ij} \equiv \del_i A_j - \del_j A_i\,.
\ee
(Note that $\dot g^{ij}$ means $\del g^{ij}/\del u$, which is the same as
$-g^{ik} g^{j\ell} \dot g_{k\ell}$.)  

   The Einstein equation $\hat R_{+-}=\Lambda \, \hat g_{+-}$ implies
\be
\ft12 H'' =\Lambda\,,
\ee
which has the immediate consequence that the $v$-dependence of
$H$ must be restricted to have the form
\be
H(u,v,x^i) = \Lambda v^2 + v\, H_1(u,x^i) + H_0(u, x^i)\,.
\label{vdep}
\ee
Thus the full system of Einstein equations 
 $\hat R_{\mu\nu}=
\Lambda \hat g_{\mu\nu}$ implies that
\bea
\nabla^2 H_0 -\ft12 F^{ij} F_{ij} - 2 A^i\del_i H_1 - H_1 \nabla^i A_i
    + 2\Lambda A^i A_i -2\nabla^i \dot A_i &&\nn\\
 + \ft12 \dot g^{ij} \dot g_{ij} + g^{ij} \ddot g_{ij} 
    +\ft12 g^{ij}\dot g_{ij} H_1&=&0\,,\label{H0eq}\\
\nabla^j F_{ij} + \del_i H_1 -2\Lambda A_i +\nabla^j\dot g_{ij} 
      -\del_i(g^{jk}\dot g_{jk}) &=&0\,,\label{nonredun}\\
\nabla^2 H_1  -2\Lambda \nabla^i A_i + \Lambda g^{ij}\dot g_{ij} 
  &=&0\,,\label{redun}\\
R_{ij}&=&\Lambda g_{ij}\,.\label{einsteq}
\eea
The coordinate $u$ plays the r\^ole of a modulus in the metrics $g_{ij}(u,x^k)$,
in the sense that (\ref{einsteq}) must hold for all values of $u$.
Note that (\ref{redun}) is redundant, since it follows by taking the
divergence of (\ref{nonredun}).  This can be seen from the fact that
under an infinitesimal deformation of the metric, the Ricci tensor
satisfies $g^{ij} \delta R_{ij}= (g_{ij} \nabla^2 - \nabla_i\nabla_j)
\delta g^{ij}$, and hence $g^{ij} \dot R_{ij} = (\nabla^i\nabla^j -
g^{ij}\nabla^2) \dot g_{ij}$.  Taking the derivative of (\ref{einsteq})
with respect to $u$, it then follows that
the divergence of (\ref{nonredun}) gives (\ref{redun}).
   
   There is a gauge symmetry of the Walker metrics (\ref{walkermet}),
which we shall discuss in the case where the Einstein equations have been
imposed.  The form of the Einstein metric is preserved under the 
transformations
\bea
&&v\longrightarrow v-f\,,\qquad A_i \longrightarrow  A_i + \del_i f\,,\nn\\
&& H_1\longrightarrow  H_1 +2\Lambda f\,,
\qquad H_0 \longrightarrow H_0 + H_1 f + \Lambda f^2 + 2\dot f\,,
\label{gauge}
\eea
where $f$ is an arbitrary function of $u$ and $x^i$.

\subsection{Reduction to a linear system for $\dot g_{ij}=0$}\label{linsys}

   If we suppose that the Einstein metric $g_{ij}$ on the transverse 
space is independent of $u$, then the remaining equations 
(\ref{H0eq})--(\ref{nonredun}) can be reduced to a linear system.  To
see this, it is convenient to use the gauge freedom (\ref{gauge}) to
impose the condition 
\be
     H_1= -\bnabla\cdot \A \label{H1con}
\ee
for all $u$.  This leaves the residual gauge freedom 
\be
A_i\longrightarrow A_i - \del_i w\,,\label{resid}
\ee
where $w(u,\x)$ satisfies the wave equation
\be
  \bnabla^2 w + 2\Lambda w=0\,.
\ee
Substituting (\ref{H1con}) into (\ref{nonredun}) shows that $A_i$ satisfies
\be
\bnabla^2 A_i + \Lambda A_i=0\,,\label{Aeq}
\ee
and so (\ref{resid}) implies that
\bea
\bnabla^2 A_i + \Lambda A_i &\longrightarrow& 
\bnabla^2 A_i + \Lambda A_i + (\bnabla^2 + \Lambda)\del_i w\nn\\
&=&\bnabla^2 A_i + \Lambda A_i + \del_i(\bnabla^2 w + 2\Lambda w)\nn\\
&=&
\bnabla^2 A_i + \Lambda A_i\,.
\eea
This means that the residual gauge transformation (\ref{resid}) could be used
to set one component of $A_i$ to zero.  

   Given any solution $A_i$ of (\ref{Aeq}), the remaining equation 
(\ref{H0eq}) becomes a Poisson equation for $H_0$, with a known source.
The solution therefore contains an arbitrary additive $u$-dependent harmonic 
function $U$ in $H_0$.  In summary, the general solution therefore depends on
$(n-2)$ arbitrary $u$-dependent 
solutions $A_i$ of the modified vector Laplace equation (\ref{Aeq})
plus the one arbitrary $u$-dependent harmonic function $U$.  The gauge freedom
(\ref{resid}) reduces this to $(n-3)$ solutions $A_i$ plus the harmonic
function $U$.

   Explicit solutions of (\ref{Aeq}) are not very easy to obtain in
general if $\Lambda\ne0$.  However, it is perhaps worth remarking that
any Killing vector field in an Einstein space automatically satisfies
(\ref{Aeq}).

\subsection{$\Lambda\ne0$ holonomy $\rm Sim(n-2)$
      solutions with $A_i=0$}\label{Aizero}

    A dramatic simplification of the linear system occurs in the case
that $A_i=0$.  This yields a very simple explicit construction of 
$n$-dimensional Einstein
metrics with proper $\rm Sim(n-2)$ holonomy, provided that $\Lambda\ne 0$.  
Thus, we can consider the following restricted class of
metrics within the Walker ansatz (\ref{walkermet}):
\be
d\hat s^2 = 2du dv + g_{ij}(\x) dx^i dx^j + [\Lambda v^2 + H_0(u,\x)] du^2\,,
\label{met3}
\ee
where $g_{ij}$ is an $(n-2)$-dimensional Einstein metric satisfying
$R_{ij}=\Lambda g_{ij}$, and $H(u,\x)$ is an arbitrary $u$-dependent 
harmonic function
\be
\nabla^i\nabla_i H_0=0\,.
\ee
The metrics (\ref{met3}) are Einstein, satisfying $R_{\mu\nu}=\Lambda 
g_{\mu\nu}$, and they have proper $\rm Sim(n-2)$ holonomy provided that
$\Lambda\ne0$, that $g_{ij}$ has maximal holonomy $SO(n-2)$, and 
that the harmonic function $H_0$ is generic.\footnote{
If, for example, $H_0$ were independent of the coordinates 
$x_i$ then the metric
$ds^2$ would simply be the direct sum of two-dimensional de Sitter or
anti-de Sitter spacetime and the $(n-2)$-dimensional Einstein metric
$g_{ij}$.  This would have holonomy $SO(1,1)\times {\rm Hol}(g_{ij})$,
which is a proper subgroup of $\rm Sim(n-2)$.}

   We can see the nature of the local holonomy explicitly by looking at
the curvature.  Choosing the natural vielbein $\hat e^a$ with
\be
\hat e^+ = dv +\ft12(\Lambda v^2 + H_0) du\,,\qquad \hat e^-=dv\,,\qquad
  \hat e^\alpha = e^\alpha\,,
\ee
we find that the non-zero components of the Riemann tensor are given by
\be
\hat R_{+-+-}= -\Lambda\,,\qquad \hat R_{-\alpha-\beta}= -\ft12\nabla_\alpha
\nabla_\beta H_0\,,\qquad \hat R_{\alpha\beta\gamma\delta}=
 R_{\alpha\beta\gamma\delta}\,.
\ee
In particular, all components of the form $\hat R_{+\alpha bc}$
are zero, which
proves that the local holonomy is either the full 
$\rm Sim(n-2)$ or a subgroup thereof.  Proper holonomy $\rm Sim(n-2)$
arises provided that $\hat R_{+-+-}\ne 0$ (and hence $\Lambda\ne 0$), 
that $g_{ij}$ has $SO(n-2)$
holonomy, and that $H_0$ is such that there are non-zero derivatives
$\nabla_\alpha\nabla_\beta H_0$.

\subsection{Previous results}

\subsubsection{Goldberg-Kerr $\Lambda=0$ metrics in $n=4$}

   Goldberg and Kerr constructed a class of 4-dimensional Ricci-flat
metrics with proper $\rm Sim(2)$ holonomy
\cite{GoldbergKerr1,GoldbergKerr2}.  They showed that 
the Weyl tensor must be of
Petrov type III, and that the metric could be cast in the form
\ben
ds ^2 = 2 du dv + dx^2 +dy ^2 + 2 \rho dx du +
\bigl (w - v \rho_x \bigr )
du^2  \label{goke}
\een
with $\rho$ and $w$ being functions of $x$, $y$ and $u$.  This therefore
corresponds to a specialisation of the discussion of Walker metrics that
we gave in section \ref{linsys}, with $n=4$ and $\Lambda=0$, and with
\be
A_1=\rho\,,\qquad A_2=0\,,\qquad H_0=w\,.
\ee
(\ie the gauge transformations have been used to set $H_1=-\bnabla\cdot A$
and $A_2=0$.)  Ricci flatness therefore implies
\bea
\rho_{xx}+ \rho_{yy}&=&0\,, \\
\omega_{xx}+ \omega_{yy}&=& 2 \rho_{-x} -2 \rho \rho_{xx}
-\bigl( \rho_x\bigr )^2 +\bigl( \rho_y\bigr )^2\,. \label{ricciflat}
\eea

   The null vector $n^\mu$ is equal to $\delta^\mu_+$, and from (\ref{nabn})
we have
\be
\nabla_\mu n_\nu = -\ft12 \rho_x\, n_\mu n_\nu\,.
\ee

  We may introduce a Majorana  representation for the four-dimensional
Clifford algebra, by defining
\bea
\Gamma^0 &=&  \pmatrix { 0 & 1 & 0 & 0 \cr  -1 & 0 & 0 & 0 \cr
  0 & 0 & 0 & -1 \cr  0 & 0 & 1 & 0 \cr }\,,\qquad
\Gamma^1  =\pmatrix { 0 & 1 & 0 & 0 \cr  1 & 0 & 0 & 0 \cr
  0 & 0 & 0 & 1 \cr  0 & 0 & 1 & 0 \cr }\,,\nn\\
\Gamma^2 &=& \pmatrix { 1 & 0 & 0 & 0 \cr  0 & -1 & 0 & 0 \cr
  0 & 0 & 1 & 0 \cr  0 & 0 & 0 & -1 \cr }\,,\qquad
\Gamma^3  =\pmatrix { 0 & 0 & 0 & 1 \cr  0 & 0 & -1 & 0 \cr
  0 & -1 & 0 & 0 \cr  1  & 0 & 0 & 0 \cr }\,.
\eea
If we then let
\ben
\epsilon= \pmatrix{\alpha \cr \beta \cr \alpha \cr \beta \cr}
\een
for constants $\alpha$ and $\beta$, we find that
\ben
\bar \epsilon \Gamma ^\mu \epsilon = 2\sqrt2 (\alpha ^2 + \beta ^2) n^ \mu
\een
and
\ben
\nabla _\mu \epsilon = -{1 \over 4} n_\mu
\pmatrix {\alpha \rho_x +\beta \rho_y \cr
\beta  \rho_x -\alpha  \rho_y \cr
\alpha \rho_x +\beta \rho_y \cr \beta  \rho_x -\alpha  \rho_y \cr } \,.
\een
In other words, we find that
\be
\nabla_\mu \epsilon = -\ft14 n_\mu\, (\rho_x - \rho_y\, \Gamma_{12})\epsilon
\,.
\ee

If one recalls that the Clifford algebra ${\rm Clif}(3,1)$ is isomorphic
with ${\rm Mat}_4({\Bbb R})$ the algebra of  $4\times$ real matrices
which is spanned by the   six antisymmetric matrices
 $C, s\Gamma_5, C \Gamma_5 \Gamma_\mu$ and the ten symmetric matrices
$C \Gamma_\mu, C \Gamma _{\mu \nu}$, \footnote{A basis may be chosen so that
$C\Gamma ^0 =1$.}, one sees that
the only  bilinears one may construct from the  commuting Majorana spinor
are
\bea
n^\mu &=& \bar \epsilon \Gamma ^\mu \epsilon \,,\\
F_{\mu \nu} &=& \bar \epsilon \Gamma _{\mu \nu}  \epsilon \,.
\eea
One finds that the only non-vanishing components of $F_{\mu \nu}$
are $F_{-i}$. Thus $F_{\mu \nu} $ is null and simple
$F_{\mu \nu} F^{\mu \nu }=0= F_{\mu \nu} \star F{\mu \nu}=0$,
with 
\ben
F_{\mu \nu} n^\mu =0\,.
\een
Since
\ben
\nabla_\rho F_{\mu \nu} = 2 \alpha  F_{\mu \nu} n_\rho
+\beta n_\rho \bigl( g_{\mu 1}
F_{\nu 2} + g_{\nu 2}  F_{\mu 1} - g_{\mu \2} F_{\nu 1} -
g_{\nu 1} F_{\mu  2}       \bigl) \,, 
\een
it follows that $F_{\mu \nu}$ is a (\lq test \rq) solution of
Maxwell's equations
\bea
\nabla _{[\rho}  F_{\mu \nu ]}&=0& \,,\\
\nabla ^\rho F_{\rho \nu }&= 0&\,.
\eea

\subsubsection{Ghanam-Thompson $\Lambda\ne0$ metric in $n=4$}

Only one $\Lambda\ne0$ solution with proper $\rm Sim(n-2)$ holonomy
has appeared previously, namely a $\rm Sim(2)$ holonomy Einstein metric
in $n=$ \cite{Ghanam}.  This metric, after correcting
typographical errors in \cite{Ghanam}, is
\bea
ds_4^2 &=& 2du dv +\fft{dx^2+dy^2}{2 x^2} + 12 xy dx du + 6(x^2-y^2) dy du\nn\\
  &&-[2v^2 + 4y(3x^2-y^2) v + (x^2+y^2)^3]du^2\,.
\eea
In fact the potential $A_i$ in this solution is pure gauge, and by
making the coordinate transformation $v\rightarrow v-y(3x^2-y^2)$ one
obtains the simpler metric form
\be
ds_4^2= 2 du dv + \fft{dx^2+dy^2}{2x^2} -
[2 v^2 + (x^2-y^2)(x^4-14 x^2 y^2 + y^4)] du^2\,.\label{ghanam2}
\ee

This four-dimensional example, as rewritten in the form
(\ref{ghanam2}), falls within the general class of $n$-dimensional
Einstein metrics that we found in
section \ref{Aizero}, for a particular choice of $H_0$ that is harmonic
in the 2-dimensional hyperbolic Einstein metric $(dx^2+dy^2)/(2x^2)$,
with $\Lambda=-2$.

\subsubsection{Brinkmann waves in $n=5$}\label{brinksec}

   The Goldberg-Kerr and Ghanam-Thompson metrics in $n=4$ dimensions
have proper $\rm Sim(2)$ holonomy.  No other Einstein metrics with proper
$\rm Sim(n-2)$ holonomy in any dimension $n$ have previously been 
exhibited.   One can of course also consider large classes of
metrics whose holonomy is a proper subgroup of $\rm Sim(n-2)$.  Our
purpose here is not to give a comprehensive review of such metrics.  
However, there is one class that we do wish to mention, since we shall
discuss them further in a later section.  These metrics are the 
Brinkmann waves in $n=5$ dimensions, which
have $\R^3$ holonomy.  

   We take the transverse metric $g_{ij}$ to be independent of $u$,
and to be that of flat Eucildean space, $g_{ij}=\delta_{ij}$.
Adopting standard the notation
of 3-dimensional Cartesian vector analysis, with $F_{ij}=\ep_{ijk} B_k$,
we have $\B=\bnabla\times\A$.  It follows from (\ref{riccigen}) that
the metric is given by (\ref{walkermet}) with
\be
\bnabla\cdot\A=0\,,\qquad \bnabla\times\A=\bnabla V\,,\qquad H= U+ \ft12 V^2\,,
\ee
where $U$ and $V$ are arbitrary $u$-dependent
harmonic functions in the transverse space, satisfying
\be
\bnabla^2 U=0\,,\qquad \bnabla^2 V=0\,.
\ee

\section{Multi-Centre Metrics}

  In this section, we first show how the five-dimensional Brinkmann wave
solutions of ${\Bbb R}^3$ holonomy
that we reviewed in section \ref{brinksec} can be used in
order to construct stationary and also time-dependent multi-black-hole
solutions upon dimensional reduction to four dimensions.  We go on to
generalise this construction by dimensionally reducing the $n=5$ 
specialisation of the ${\rm Sim}(n-2)$
holonomy solutions  with cosmological constant that we obtained in 
section \ref{Aizero}.

\subsection{Time-independent Kaluza-Klein black holes}\label{tindep}

If $U$ and $V$ are chosen to be independent of $u$, and if $H>0$, 
then $\del/\del u$ is a spacelike Killing vector field that can be used for
performing a Kaluza-Klein reduction, using the standard formula
\be
ds_5^2 = e^{-2\phi/\sqrt3}\, ds_4^2 + e^{4\phi/\sqrt3}\, (du + 2B)^2\,,
\label{kk54}
\ee
in which the lower-dimensional metric $ds_4^2$ is in the Einstein
conformal gauge.
This leads to a stationary 4-dimensional
metric in which the coordinate $v$ plays the r\^ole of time.  The 
four-dimensional metric is
\be
ds_4^2 = - H^{-1/2} (dv+A)^2 + H^{1/2} dx^i dx^i\,,
\ee
and the Kaluza-Klein vector and scalar are given by
\be
B= \fft{1}{2H}\, (dv + A)\,,\qquad e^{4\phi/\sqrt3} = H\,.
\ee
In the conventions we are using, the lower-dimensional Lagrangian is
given by
\be
{\cal L} =\sqrt{-g}\Big( \ft14 R -\ft12 (\del\phi)^2 - 
                           \ft14 e^{2\sqrt3\, \phi} G^2\Big)\,,
\label{d4lag}
\ee
where $G=dB$.

   If we choose 
\be
V=0\,, \qquad H= 1 + \sum_{a=1}^N \fft{4M_a}{| \x-\x_a |}\,,
\ee
we obtain the metric describing $N$ extremal Kaluza-Klein black holes
with masses $M_a$, charges $2M_a$, and scalar charges $\sqrt3 M_a$ in
equilibrium \cite{GibbonsNP}.  

   Adding a harmonic function $V$ of the form
\be
V= \sum_{a=1}^N \fft{N_a}{|\x - \x_a|}
\ee
would maintain the equilibrium by endowing these objects with 
NUT charges proportional to $N_a$.  However, the resulting metrics would
not be asymptotically flat.  To obtain asymptotically flat metrics with
angular momentum, $V$ could be chosen to have the form of a sum of
dipoles, the angular momenta being proportional to the dipole moments.

\subsection{Time-dependent Kaluza-Klein black holes}\label{tdep}

   To obtain time-dependent solutions in four dimensions, we choose
$U$ and $V$ so that the five-dimensional metric is invariant under the
$SO(1,1)$ boost action $u\rightarrow \lambda u$, $v\rightarrow v/\lambda$.
To achieve this we introduce new coordinates $t$ and $z$ defined by
\be
u= -\fft{2}{h}\, e^{-hz/2}\,,\qquad v= t e^{hz/2}\,,\label{uvredef}
\ee
where $h$ is an arbitrary constant, and we take $U$, $V$ and $A_i$ to have
the specific $u$-dependences
\be
U(u,\x)= \fft{4\wtd U(\x)}{h^2 u^2}\,,\qquad
V(u,\x)= \fft{2\wtd V(\x)}{h u}\,,\qquad
 A_i(u,\x)= -\fft{2 \wtd A_i(\x)}{hu}\,.
\ee
These $u$-dependences, which also imply that $H(u,\x)= 4\wtd H(\x)/(h^2 u^2)$
with $\wtd H = \wtd U + \ft12 \wtd V^2$, ensure that the five-dimensional
metric is indeed boost invariant.  It takes the form
\be
ds_5^2 = -(\wtd H + ht)^{-1} (dt + \wtd A)^2 + dx^i dx^i +
  (\wtd H+h t)\Big( dz + \fft{dt+\wtd A}{\wtd H + h t}\Big)^2\,,
\ee
and thus it reduces to give
\bea
ds_4^2 &=& -(\wtd H + ht)^{-1/2} (dt + \wtd A)^2 + 
                   (\wtd H + ht)^{1/2} dx^i dx^i\,,\nn\\
B &=& \ft12 (\wtd H + ht)^{-1}\, (dt+\wtd A)\,,
\qquad e^{4\phi/\sqrt3} = \wtd H + ht
\eea
in four dimensions.

   These time-dependent solutions have a similar interpretation to the
metrics introduced in \cite{GibbonsLuPope}.  One has extreme Kaluza-Klein
black holes moving in a background $k=0$ FLRW universe dominated 
by a massless scalar field, and hence with scale factor 
$a(\tau)\propto \tau^{1/3}$.

\subsection{Kaluza-Klein monopoles}

   The four-dimensional field equations admit electric-magnetic duality,
whereby
\be
\phi\longrightarrow -\phi\,,\qquad G_{\mu\nu}\longrightarrow 
 \overline {G}_{\mu\nu} =e^{2\sqrt3 \phi}\, {*G_{\mu\nu}}\,,\qquad
g_{\mu\nu}\longrightarrow g_{\mu\nu}\,.
\ee
One finds that
\be
\overline{G}=d\overline{B} = \ft12 ({*_3\wtd F})\wedge (dt+\wtd A) -
     \ft12 {*_3 d} \wtd H + \ft12 h {*_3 \wtd A}\,.
\ee
Lifting back to five dimensions, this gives
\be
ds_5^2 = -(dt+\wtd A)^2 + (\wtd H+ ht) dx^i dx^i + 
            (\wtd H+ht)(dz + 2 \overline{B})^2\,.
\ee
If $h=0$ and $\wtd V=0$, we obtain \cite{GibbonsNP} the static multi-monopole
Kaluza-Klein 5-metric, which is the direct sum of a four-dimensional 
gravitational multi-instanton \cite{Hawking,GibbonsHawking} and time.
This has holonomy $SU(2)$ rather than the holonomy $\R^3$ that we started
with.  The more general cases for which $h$ and $\wtd V$ are non-vanishing
represent moving Kaluza-Klein monopoles which may have angular momenta
and NUT charges.

\subsection{Time-dependent cosmological black holes}

   If a spacelike Killing vector exists, then a dimensional reduction
is still possible even if $\Lambda\ne0$.   
The four-dimensional Lagrangian (\ref{d4lag}) then contains an additional
Liouville potential for the scalar $\phi$:
\be
{\cal L} =\sqrt{-g}\Big( \ft14 R -\ft12 (\del\phi)^2 -
                           \ft14 e^{2\sqrt3\, \phi} G^2 -W(\phi)\Big)\,,
\qquad W(\phi)= \ft34\Lambda e^{-2\phi/\sqrt3}\,.
\label{d4lag2}
\ee

    We may take as our starting point the $u$-independent 
5-dimensional metrics of the type we considered in section \ref{Aizero},  
\ie where $A_i=0$ and $H_0$ can be an arbitrary harmonic function on
the transverse space.  Since the metric $g_{ij}$ is 3-dimensional and
Einstein with $\Lambda\ne0$, its universal cover must be either 
$S^3$ or the hyperbolic space $H^3$, depending on whether $\Lambda$ 
is positive or negative.  Using (\ref{kk54}), the five-dimensional
Einstein metric reduces to give the
four-dimensional solution
\bea
ds_4^2 &=& -\fft{dv^2}{(\Lambda v^2 + H_0)^{1/2}} +
   (\Lambda v^2 + H_0)^{1/2}\, g_{ij} dx^i dx^j\,,\nn\\
\phi &=& \ft{\sqrt3}{4} \log(\Lambda v^2 + H_0)\,,\qquad
B=\fft{dv}{2(\Lambda v^2 + H0)}\,.\label{cossol}
\eea
These metrics represent extremal charged black holes moving in a 
cosmological background.  
   
    Note that this reduction of $u$-independent five-dimensional 
solutions has yielded {\sl time-dependent} metrics in four dimensions, in
contrast to the reduction of $u$-independent Brinkmann solutions we 
performed in section (\ref{tindep}).

    It is convenient to define a rescaled transverse metric 
$\hat g_{ij}= \ft12|\Lambda| g_{ij}$, so that the radius of curvature 
is $k=\pm 1$.  If $H_0$ is taken to be independent of $x^i$ (\ie
$H_0$ is a constant), then the metric (\ref{cossol}) 
can be cast into the standard FLRW form by introducing a proper-time
coordinate $\tau$ according to
\be
d\tau = \fft{dv}{(\Lambda v^2 + H_0)^{1/4}}\,,
\ee
and defining the scale factor 
\be
a(\tau)= \Big(\fft{2}{|\Lambda|}\Big)^{1/2}\,
          (\Lambda v^2 + H_0)^{1/4}\,.
\ee
The metric then be written as
\be
ds_4^2 = -d\tau^2 + a^2(\tau)\, \hat g_{ij}\, dx^i dx^j\,.
\ee
The scale factor $a(\tau)$ may be seen to satisfy the Friedman equation
\be
\Big(\fft{\dot a}{a}\Big)^2 + \fft{k}{a^2} =
      \ft23 \Big( \ft12 \dot\phi^2 + W(\phi)\Big)\,,\label{friedman}
\ee
and $\phi$ satisfies
\be
\fft1{a^3}\, \fft{d(a^3\dot\phi)}{d\tau} + \fft{dW(\phi)}{d\phi}=0\,.
\ee
Note that here, we are using a dot to denote a derivative with respect
to $\tau$.  In fact the solution has the property that the derivative
and the non-derivative terms in (\ref{friedman}) are separately equal,
\be
\Big(\fft{\dot a}{a}\Big)^2 = \ft13 \dot\phi^2\,,\qquad \fft{k}{a^2} =
    \ft23 W(\phi)\,. \label{first}
\ee
The equations of motion for a FLRW model of this kind can be derived from
the Lagrangian
\ben
L=a^3 \Bigl [N^{-1} \Bigl
( \half \dot \phi^  2 - {3 \over 2} {\dot a ^2 \over a^2 } \Bigr )  -N \Bigl
( W(\phi )
+ { 3 k \over a^2 } \Bigr )  \Bigr ]\,,
\een
where the lapse $N$ is a Lagrange  multiplier 
enforcing the constraint that the associated  Hamiltonian $\cal H $ 
vanishes \ben
{\cal H} = a^3 \Bigl [ \half \dot \phi ^ 2 - {3 \over 2} {\dot a ^2 \over a^2 }
+  W(\phi )
- { 3 k \over a^2 }  \Bigr ] =0\,.
\een
The vanishing of the Hamiltonian $\cal H $ is equivalent to the
Friedman equation (\ref{friedman}). The first order equations
(\ref{first}) resemble in some ways  Bogomol'nyi equations
but  they appear not to be derivable from a super-potential.   
  
 As in our previous discussion of time-dependent solutions in 
section \ref{tdep}, instead
of starting from $u$-independent 5-dimensional solutions and reducing on
$\del/\del u$ we may alternatively start from 5-dimensional solutions
with the very specific $u$-dependence that ensures boost invariance
under $u\rightarrow \lambda u$, $v\rightarrow v/\lambda$.  Again, the
dimensional reduction is then performed by defining new coordinates 
as in (\ref{uvredef}), and then reducing on the spacelike Killing 
vector $\del/\del z$.  Starting again with the 5-dimensional metrics
considered in section \ref{Aizero}, but now taking the harmonic function to
have the form
\be
H_0(u,\x) = \fft{4 \wtd H_0(\x)}{h^2 u^2}\,,
\ee
we obtain the metric
\be
ds_4^2 = -\fft{dt^2}{(\Lambda t^2 + ht + \wtd H_0)^{1/2}} +
      (\Lambda t^2 + ht + \wtd H_0)^{1/2}\, g_{ij} dx^i dx^j
\ee
after dimensional reduction.  This is in fact equivalent to the previous 
metric in (\ref{cossol}), as can be seen by performing the redefinitions
\be
t\longrightarrow v -\fft{h}{2\Lambda}\,,\qquad \wtd H_0 \longrightarrow
   H_0 + \fft{h^2}{4\Lambda^2}\,.
\ee

   In contrast to the situation for our dimensional reductions of Brinkmann 
waves, where the reductions on $\del/\del u$ in section \ref{tindep}
gave time-independent four-dimensional solutions, whilst reductions on
the boost Killing vector $\del/\del z$ in section \ref{tdep} gave 
time-dependent solutions, we see that when $\Lambda$ is non-zero both the
$\del/\del u$ and $\del/\del z$ reductions give time-dependent solutions 
in four dimensions, and in fact the two reduction schemes give equivalent
such solutions.

   Another difference between the Brinkmann reductions and the cosmological
reductions is that the former give rise to multi-centre black holes in a
$k=0$ FLRW background, whilst the latter give rise to multi-centre black
holes in $k=\pm 1$ FLRW backgrounds.

\section{Solutions with $u$-Dependent Transverse Metrics}

   So far, we have considered only situations where the transverse metric
$g_{ij}$ is independent of $u$.  However, we could take for $g_{ij}$ an
arbitrary curve in the space of Einstein metrics.  If we were to do so,
our general reduction of the remaining equations to a linear system is 
still possible, as long as the extra terms involving $\dot g_{ij}$ are
included as sources.  

    The curve of Einstein metrics may or may not be among metrics that 
are related by diffeomorphisms.  In the former case, it seems likely 
that by means of a coordinate transformation one may pass to the case 
where the transverse metric is independent of $u$.  A situation in which
this is definitely the case is when the metric $g_{ij}$ is flat, with
\be
g_{ij}(u,\x) = \gamma_{ij}(u)\,,\qquad  A_i=0\,,\qquad H_1=0\,,
\ee
and so
\be 
ds^2 = 2 du dv + \gamma_{ij}(u)\, dx^i dx^j + H_0(u,\x)\, du^2\,.
\label{umet}
\ee

    We pass to new  coordinates $\tilde x ^m$, $\tilde u$ and $\tilde v$,
given by
\ben
x^i= P^i\,_m (u) \tilde x ^m \,,\qquad 
v= \tilde v - \tilde x^m A_{mn}(u) \tilde x ^n\,,\qquad u=\tilde u\,,
\label{tildes}
\een
where $P^i{}_m$ is chosen such that
\ben
\g_{ij}(u) P^i\,_m (u)  P^j\,_{n} (u) = \delta_{mn} \,.\label{Pdef}
\een
Furthermore,  $A_{mn}(u)$, which is symmetric, will be chosen to eliminate 
the resulting
$d\tilde x^m d\tilde u$ cross terms in the metric. 

    In what follows it will prove convenient to adopt a matrix notation and 
write this as
\ben
P^t \g P =1\,,\label{gamma}
\een
where $^t$ denotes transpose. This can equivalently be written as
\be
\gamma = (P P^t)^{-1}\,,\label{gamma2}
\ee
from which it is clear that the choice of $P$ is arbitrary up to an 
$O(n-2)$ transformation $U$:
\be
P'= PU\,.\label{PU}
\ee
We also define a matrix
\ben
B= P^{-1} \dot P \,,\label{Bdef}
\een
where in this section $\dot P $ denotes differentiation with respect
to $u$.  

   The $d\tilde x^m d\tilde u$ cross terms coming from the metric
(\ref{umet}) will be absent if $A$ can be chosen so that $2A = 
P^t\gamma \dot P$, which, using (\ref{gamma}) and then (\ref{Bdef}), means
\be
2A = P^{-1} \dot P = B\,.\label{Adef}
\ee
In general, $B=P^{-1}\dot P$ is not symmetric, unlike $A$.  However, we can
use the freedom to perform the $u$-dependent  $O(n-2)$ transformation 
(\ref{PU}) in order to find a suitable $P'$ for which $B'\equiv {P'}^{-1}
\dot P'$ {\it is} symmetric.  Specifically, $U$ should be chosen so that
\be
B^t -B = 2 \dot U U^{-1}\,.
\ee
This always admits a solution for $U$, since there are $\ft12 (n-2)(n-3)$ 
first-order equations for $\ft12 (n-2)(n-3)$ unknown functions.  Having
seen that a gauge can be achieved where $B'$ is symmetric, we may now
drop the prime and assume that such a symmetric $B$ exists.

   Having eliminated the $d\tilde x^m d\tilde u$ cross terms in the
metric by choosing $A$ to satisfy (\ref{Adef}), the metric (\ref{umet})
in the tilded coordinates takes the form
\be
ds^2 = 2 d\tilde u d\tilde v + d\tilde x^m d\tilde x^m + \widetilde H\, 
   d\tilde u^2\,.\label{tildemet}
\ee
where
\be
\widetilde H = H_0 + \tilde x^m \widetilde K_{mn} \tilde x^n\,,\label{tHdef}
\ee
and 
\be
\widetilde K = \dot P^t \gamma \dot P - 2 \dot A\,.
\ee
 Using (\ref{gamma}) and (\ref{Adef}), we have $ \dot P^t \gamma \dot P=
B^t P^t\gamma PB=B^2$, and hence
\be
\widetilde K = B^2 -\dot B\,.
\ee

   From (\ref{H0eq}), the function $H_0$ satisfies
\be
\gamma^{ij} \del_i\del_j H_0 + \ft12 \dot\gamma^{ij} \dot\gamma_{ij} + 
   \gamma^{ij} \ddot\gamma_{ij}=0\label{H0orig}
\ee
in the original coordinate system.  Using (\ref{gamma2}), (\ref{Bdef}) and
(\ref{Adef}), it can be seen that
\be
\dot \gamma^{-1} \dot \gamma =  -4 P B^2 P^{-1}\,,\qquad 
  \gamma^{-1} \ddot \gamma = 2P(2B^2 -\dot B)P^{-1}\,,
\ee
and so (\ref{H0orig}) becomes
\be
0=\gamma^{ij} \del_i\del_j H_0 + 2\tr(B^2 -\dot B)=
   \gamma^{ij} \del_i\del_j H_0 + 2 \tr \widetilde K\,.
\ee
Transforming to the tilded coordinates (\ref{tildes}), we see that 
the function $\widetilde H$ in the transformed metric (\ref{tildemet})
must simply obey the harmonic equation
\be
\tilde\del_m \tilde\del_m \widetilde H =0\,.
\ee

\subsection{The reverse transformation: Rosen waves}

We may obviously carry out the previous steps in reverse.  
Suppose that
\ben
ds ^2 = 2 d \tilde u d \tilde v + \tilde H(\tilde u , \tilde x ^i) 
            d\tilde u^2  + d \tilde x^i d \tilde x^i \,.
\een

Let 
\ben
\tilde x =M (u)  x\,\qquad \tilde v=v + x^t N(u)  x\,,\qquad \tilde u =u\,. 
\een
where without loss of generality $N(u)$ can be assumed to be symmetric.
The transformed  metric is
\ben
ds ^2 = 2 du dv + \g_{ij}(u) dx^i dx^j + H(u,x^i) du ^2 \,, 
\een
where
\bea
\g_{ij}(u)&=& M^t M \,,\\
 H &=& \tilde H +  x ^t
 \Bigl (  2 \dot N + \dot M^t \dot M  \Bigr ) x \,,\label{dotN}
\eea
and the cross terms between $dx^i$ and $du$ are eliminated if $N$ and $M$
are such that
\ben
2N + M^t \dot M   = 0\,. \label{N}
\een
Thus to remove the cross terms, it must be that $M^t\dot M$ is symmetric.
However, in  the notation of the previous section,
\ben
M=P^{-1}\,,\qquad N=M^t A M\,, 
\een
and
\ben
M^t\dot M= - M^t B M \,.
\een
Thus we may use the same gauge freedom 
as before to arrange that $M^t\dot M$ is symmetric.

Differentiation  of (\ref{N}) and elimination
of $ \dot N$ from (\ref{dotN}) gives
\ben
H= \tilde H -   x^t (M^t \ddot M)  x 
\,.
\een

Now suppose that 
\ben
\tilde H = \tilde x ^t \tilde  K \tilde x = x^t M^t K M x \,,
\een 
with $K=L^t, \tilde K = \tilde K^t$.
We can then set $H=0$ by solving
\ben
 \ddot M = K M \,. 
\een
Given $ L$, this is a linear  second order differential equation for $M$
which can always be solved.
The result is that in the $x,u,v$ coordinates, the metric
is independent of $x$, and the metric $\gamma_{ij} $ satisfies 
the single condition vacuum field equation

\be
\ft12 \dot\gamma^{ij} \dot\gamma_{ij} + 
   \gamma^{ij} \ddot\gamma_{ij}=0\,,.\label{field}
\ee
Such solutions are called plane gravitational waves 

The $1+2(n-2)$  vector fields
\ben
V= \pa_v\,,\qquad  P_i = \pa_i 
\een
and 
\ben
U_i = x_i \pa_v  + u \pa_i 
\een
generate isometries \cite{BondiPiraniRobinson,GibbonsRuback}. Since, 
in general, $\gamma_{ij}$ will have no isometries and so
we omit
\ben
L_{ij}= x_i\pa_j-x_j \pa_j \,.
\een
As a result the symmetry of the Rosen waves is 
the  subgroup of the Carroll group \cite{Leblond,Sen} 
(itself generated by $U_i, P_i, V, L_{ij}$) 
which omits the rotations $L_{ij}$.

The reason for the name Carroll group
is  explained in the appendix.

\subsection{Time dependence of curved transverse metrics}
If
\ben
ds ^2 = 2 du \Bigl(  dv + A_i(u,x)  dx ^i \Bigl) + g_{ij}(u,x) dx ^i dx ^j
+H (u,v, x)  du^2 \,,
\een
then under the coordinate transformation
\ben
v=\tv + F(u,\tx ) \,,\qquad x^i= a^i( u,\t, x) \,,
\een 
we shall have
\bea
H &\rightarrow & H + 2 \dot F + g_{ij} \dot a^i \dot a^j + 2 A_i \dot a^i\,,
\\
A _i &\rightarrow & A_r {\pa x^r \over \pa \tx ^i } + {\p F \over \p \tx
  ^i} 
    + g_{rs} \dot a^r {\p x^s \over \p \tx ^i}\,,\\ 
g_{ij} &\rightarrow & g_{rs} {\p a^r \over \p \tx ^i} {\p a^s \over 
\p  \tx ^i} \,.
\eea

   Thus if $g_{ij}(u,x)$ is, for each fixed $u$, a coordinate
transformation of a fixed $u$-independent metric, then we can choose 
coordinates so that
$g_{ij}$ is independent of $u$, by choosing an appropriate
$a^i(u,\tx)$.
If there is only one Einstein metric, with fixed $\Lambda$, 
up to diffeomorphisms on the base manifold, then this is always the case. 

If, however, the space of Einstein metrics on the base manifold has
non-trivial moduli,
$g_{ij}(u,x)$ could pass along a path of non-diffeomorphic metrics
and the time-dependence cannot be eliminated in this way.

Indeed either the curve of metrics can be chosen
arbitrarily, and for each $u$ we can solve for $A_i$, $H_1$ and $H_0$,
or we can set, for example, $A_i=0=H_1=H_0$ and try to find conditions on 
$g_{ij}(u.x)$. These might be consistent time-dependent Calabi-Yau's
for example. 

A separate question, discussed  earlier in the case of flat metrics,
 is whether the cross-term $A_i$ can be eliminated or kept non-zero by
 a suitable choice of $F(u,\tx ) $.

\section{Conclusion}

In this paper we have shown that the problem of finding the general
$n$-dimensional Lorentzian  Einstein metric with ${\rm Sim}(n-2)$ 
may be reduced to solving a set of linear equations
on an $(n-2)$-dimensional transverse metric which itself is a, possibly
time dependent,  Einstein metric.
We have also shown how the metric may be used to construct 
new four-dimensional  multi-centre metrics in which  
extreme Kaluza-Klein black holes move in a background  
F-L-R-W metric with non-vanishing space curvature
coupled to  a scalar field 
with a Liouville potential.

We believe that the general ${\rm Sim}(n-2)$ holonomy metrics
we have constructed in this paper will find various other applications
in M-theory and string theory in the future.

\section*{Acknowledgements}

We should like to thank Joaquim Gomis, Graham Hall, Sigbjorn Hervik
and Maciej Dunajski for helpful discussions.   The research of 
C.N.P. is supported in part by DOE grant DE-FG03-95ER40917.

\newpage

\appendix

\section{${\rm Sim} (n-2)$, $ {\rm ISim} (n- 2)$ 
and the Carroll Group}

In this appendix we recall some facts about the Carroll
group explained 
in \cite{DuvalGibbonsHorvathy} and relate them to the groups
${\rm Sim} (n-2)$, $ {\rm ISim} (n- 2)$.

The Carroll group, and the Galilei group 
are both kinematic groups of a spacetime in the sense of \cite{Bacry,Nuyts}  
and both may be regarded as the symmetry group
of a structure in a Lorentzian spacetime with  one  higher dimensions.

We start with a construction of the  Galilei group. The basic idea is 
to start with  flat Minkowski spacetime ${\Bbb E} ^{n,1}$ whose metric
written in
double null coordinates  $(u,v,x^i)$, $i=1,2,\dots, n-1$,  is
\ben
ds^2 =- 2 dudv +  dx^i d x^i. 
\een  

The Lie algebra of the Poincar\'e group $\frak{e}(n,1)$ is spanned by
the Killing vector fields generating the Lie algebra of the Euclidean
group $\frak{e}(n-1)$, translations and rotations   
\ben
P_i = \partial _i \qquad  L_{ij}= x_i \partial _j - x_j \partial _i,
\een
two null translations and one boost
\ben
U=\partial _u, \qquad V= \partial _v \qquad N= u \partial _u -v \partial _v, 
\een
and two further sets of boosts

\ben
U_i= u\partial _i + x _i \partial _v \qquad V_i=  v \partial _i + x_i
\partial _u.
\een
There is an obvious symmetry under inter-changing $u$ and $v$ induced 
by reflection in the timelike $(n-1)$- plane $u=v$. 

To obtain the Bargmann group, the central extension of the
Galilei group, 
we ask for the subgroup which 
commutes with the null translation generated by $V=\partial _v$. 
This is generated by $\{ P_i, L_{ij}, U, V, U_i,   \}$. The Galilei
group is obtained by taking the quotient by the null translation group
$\Bbb R $ generated by $V$.
It is easy to see
that the Galilei group acts on the quotient  ${\Bbb E}^{n,1} /{\Bbb
 R}$ or light-like shadow,
 which may be identified with a  Newton-Cartan spacetime $M^n $, 
the coordinate $u$ playing the role of Newtonian absolute time.
The generators $U_i$
are Galilean boosts. Because 
\ben
[P_i, U_j ]= \delta_{ij} V, \label{Heisenberg}
\een
they commute with
spatial translations (modulo $V$) 
but not with time translations
\ben
[U, U_i]=P_i.
\een  

One  may regard   this construction in terms of a Kaluza-Klein type reduction
in which one  think of ${\Bbb E} ^{n,1}$ as a fibre bundle
with projection map \ben \pi: {\Bbb E} ^{n,1} 
\rightarrow M^n \label{submersion}\een 
given by $ (u,v,x^i)  \rightarrow (u, x^i)$. 
However in contrast to the usual case, the fibres are lightlike.
Using the map $\pi $ one may push forward the Minkowski  co-metric
on ${\Bbb E} ^{n-1}$ down to the Newton--Cartan
${\Bbb E} _{n-1,0}$ spacetime to give the  degenerate co-metric.
More about lightlike reduction may be found in \cite{JuliaNicolai,Minguzzi1}.
For an interesting application  of 
the inverse process, lightlike oxidation, see \cite{Minguzzi2}. 

To obtain the Carroll group, we ask instead for the subgroup  
of the Poincar\'e group which leaves invariant the null hyperplane
$u={\rm constant}=0$. This is generated by $\{P_i, L_{ij}, V, U_i ,N\}$.
 To obtain 
the Carroll group we quotient by the boots $N$ . 
Now the null coordinate $v$ plays the role of time. The Carollian
boosts
$U_i$ commute with time translation
\ben
[V, U_i]=0,
\een
 but by (\ref{Heisenberg}) they  no longer commute
with spatial translations $P_i$. In fact one obtains
a  Heisenberg sub-algebra
with the time translations being central.          
Note that the boots $N$, which generate
\ben
v \rightarrow \lambda v\,,\qquad u \rightarrow \lambda ^{-1} u
\een
act as time dilations 
\ben
t \rightarrow \lambda  t \,,\qquad x_i \rightarrow x_i\,. 
\een
From an algebraic point of view the Carrol and Galilei groups
differ only in the choice of generator of time translations: one picks
either $V$ or $U$.

One may think of the null hyperplane $u={\rm constant}$ as
the image under the embedding map \ben x: M^n \rightarrow {\Bbb E}
^{n,1} \label{immersion}, \een such that $(v,x^i) \rightarrow ({\rm
  constant},v,x^i)$,    
of a Carollian spacetime time. The pull back of the Minkowski metric
gives the degenerate Carrollian metric .
Thus the duality relating the cases  is between an  immersion
$x$ (\ref{immersion})
and a submersion $\pi$  (\ref{submersion}) and interchanges 
domain and range. 
 
If we retain the generators
$U_i, N, L_{ij}$ we obtain an $\half(n^2 -3n +4)$- dimensional
subgroup of the Lorentz group $SO(n-1,1)$ invariant
and which normalises  the lightlike  vector field $V$,
\ben
[N,V] = V\,.
\een
Since 
\ben
[U_i,U_j]=0\,,\qquad  
[N, L_{ij}]=0\,,\qquad [N, U_i] = U_i\,,  
\een
This group is isomorphic to the
Euclidean group $E(n-2)$ augmented with homotheties,
and is thus called  $Sim(n-2)$. Together with the translations 
$p_i,U,V$    one obtains an $\half(n^2 -n +4)$- dimensional
subgroup of the Poincar\'e group  called ${\rm ISim} (n-2)$ 
which, in the case $n=4$   is the basis
of Very Special Relativity \cite{CohenGlashow,GibbonsGomisPope} which 
is used as   model of broken Lorentz-invariance with no invariant tensor
fields, called in this context spurion fields.


\newpage

\begin{thebibliography}{99}

\bibitem{Scherk} J. Scherk, 
{\it Antigravity: A crazy idea?},
Phys.Lett. {\bf B  88}, 265 (1979).
  
\bibitem{HartleHawking} J.B. Hartle and S.W. Hawking,
{\it Solutions of the Einstein-Maxwell equations with many black holes},
Commun. Math. Phys. {\bf 26} (1972) 87.

\bibitem{GibbonsTownsend} G.W. Gibbons and P.K. Townsend,
{\it Vacuum interpolation in supergravity via super p-branes},
Phys.Rev. Lett.   {\bf 71} (1993) 3754, hep-th/9307049.

\bibitem{Hawking} S.W. Hawking, 
{\it Gravitational instantons}, Phys. Lett. {\bf A  60} (1977) 81.

\bibitem{GibbonsHawking} G.W. Gibbons and S.W. Hawking,
{\it Gravitational multi-instantons,}
Phys  Lett.  {\bf B 78} (1978) 430.

\bibitem{GrossPerry} D.J. Gross and M.J. Perry,
{\it Magnetic monopoles in Kaluza-Klein theories},
Nucl. Phys.  {\bf B  226} (1983) 29.

\bibitem{Sorkin} R.D. Sorkin, 
{\it Kaluza-Klein monopole},
Phys. Rev. Lett.  {\bf 51} (1983) 87.

\bibitem{GibbonsHorowitzTownsend} G.W. Gibbons, G.T. Horowitz and P.K. 
Townsend,
{\it Higher dimensional resolution of dilatonic black hole singularities},
Class. Quant. Grav. {\bf 12} (1995) 297, hep-th/9410073.

\bibitem{Welch} D.L. Welch,
{\it On the smoothness of the horizons of multi-black hole solutions},
Phys. Rev.  {\bf D 52} (1995) 985, hep-th/9502146.

\bibitem{CandlishReall} G.N. Candlish and H.S. Reall,
{\it On the smoothness of static multi-black hole solutions of
higher-dimensional Einstein-Maxwell theory},
arXiv:0707.4420[gr-qc].

\bibitem{GibbonsRuback} G.W. Gibbons and P.J. Ruback,
{\it Classical gravitons and their stability in higher dimensions},
Phys.Lett. {\bf B 171}, 390 (1986).

\bibitem{CsakiShirman}
  C.~Csaki and Y.~Shirman,
Brane junctions in the Randall-Sundrum scenario,
{\it  Phys. Rev.} {\bf D  61} (2000) 024008, hep-th/9908186.


\bibitem{KastorTraschen} D. Kastor and J.H. Traschen,
{\it Cosmological multi-black-hole solutions},
Phys. Rev.  {\bf D  47} (1993) 5370, hep-th/9212035.

\bibitem{MakiShiraishi} T. Maki and K. Shiraishi,
{\it Multi-black-hole solutions in cosmological Einstein-Maxwell dilaton},
Class. Quant. Grav. {\bf 10} (1993) 2171.

\bibitem{GibbonsLuPope} G.W. Gibbons, H. L\"u and C.N. Pope,
 {\it Brane worlds in collision},
 Phys. Rev. Lett.  {\bf 94} (2005) 131602, hep-th/0501117.

\bibitem{ChenChongGibbonsLuPope} W. Chen, Z.W. Chong, G.W. Gibbons, 
H. L\"u and C.N. Pope,
{\it Horava-Witten stability: Eppur si muove},
Nucl. Phys.  {\bf B 732} (2006) 118, hep-th/0502077.

\bibitem{GibbonsMunich} G.W. Gibbons,
{\it Supersymmetric soliton states in extended supergravity theories},
in Muenchen 1981, Proceedings, {\it Unified Theories Of Elementary 
Particles}, 145-151.

\bibitem{GibbonsTrieste} G.W. Gibbons,
{\it The Bogomolny inequality for Einstein-Maxwell theory},
in Trieste 1981, Proceedings, {\it Monopoles In Quantum Field Theory},
 137-138.

\bibitem{GibbonsLondon} G.W. Gibbons,
{\it The multiplet structure of solitons in the $O(2)$ supergravity theory},
in  Proccedings...

\bibitem{GibbonsHull} G.W. Gibbons and C.M. Hull,
{\it A Bogomolny bound for general relativity and solitons in $N=2$
supergravity}, 
Phys.Lett  {\bf B109} (1982) 190.

\bibitem{Freedman} D.Z. Freedman, G.W. Gibbons and M. Schnabl,
{\it Matrix cosmology}, 
AIP Conf. Proc.  {\bf 743}, 286 (2005), hep-th/0411119.

\bibitem{Brinkmann1} M.W. Brinkmann, 
{\it On Riemann spaces conformal to Euclidean space}, 
Proc. Natl. Acad. Sci. U.S. {\bf 9}, 1 (1923).

\bibitem{Brinkmann2} M.W. Brinkmann, 
{\it On Riemann spaces conformal to Einstein spaces},
Proc. Natl. Acad. Sci. U.S. {\bf 9}, 172 (1923).

\bibitem{Brinkmann3} M.W. Brinkmann,
{\it Einstein spaces which are mapped conformally on each other},
Math. Ann. {\bf  94}, 119 (1925).

\bibitem{Duval} C. Duval, G. Burdet, H.P. Kunzle and M. Perrin,
{\it Bargmann structures and Newton-Cartan theory,}
Phys. Rev. {\bf D31}, 1841 (1985).

\bibitem{DuvalGibbonsHorvathy} C. Duval, G.W. Gibbons and P. Horvathy,
{\it Celestial mechanics, conformal structures, and gravitational waves},
Phys. Rev. {\bf  D43}, 3907 (1991), hep-th/0512188.

\bibitem{JuliaNicolai} B. Julia and H. Nicolai,
{\it Null Killing vector dimensional reduction and Galilean geometrodynamics},
Nucl. Phys.  {\bf B  439}, 291 (1995), hep-th/9412002.

\bibitem{GibbonsNP} G.W. Gibbons,
{\it Antigravitating black hole solitons with scalar hair in $N=4$ 
supergravity},
Nucl. Phys. {\bf B207}, 337 (1982).

\bibitem{Schell} J.F. Schell, 
{\it Classification of four-dimensional Riemannian spaces}, 
J. Math. Phys. {\bf 2}, 202 (1961).

\bibitem{Ghanam} R. Ghanam and G. Thompson, 
{\it Two special metrics with $R_{14}$-type holonomy}, 
Class. Quant. Grav. {\bf 18}, 2007 (2001). 

\bibitem{GoldbergKerr1} R.P. Kerr and J.N. Goldberg
{\it Some applications of the infinitesimal-holonomy group to the Petrov
classification of Einstein spaces}, 
J. Math. Phys. {\bf 2}, 327 (1961).

\bibitem{GoldbergKerr2} R.P. Kerr and J.N. Goldberg
{\it Einstein spaces with four-parameter holonomy groups},
J. Math. Phys. {\bf 2}, 332 (1961).

\bm{exact} H. Stephani, D. Kramer, M. MacCallum, C. Hoenselaers and
E. Herlt, {\it Exact solutions to Einstein's field equations} (Second
edition), (CUP 2003).

\bibitem{Ortaggio} M. Ortaggio, 
{\it Higher dimensional spacetimes with a geodesic, shearfree,
twistfree and expanding null congruence}, 
in Proceedings of the XVII SIGRAV Conference, Turin, September 4--7, 2006,
gr-qc/0701036. 

\bibitem{Walker} A.G. Walker, 
{\it Canonical form for a Riemannian space
with a parallel field of null planes},
Quart. J. Math., Oxford (2) {\bf 1}, 69 (1950).

\bibitem{BondiPiraniRobinson} H. Bondi, F.A.E.  Pirani and I. Robinson,  
{\it Gravitational waves in general relativity III: Exact plane waves},
Proc. Roy. Soc. London Ser. {\bf  A 251}, 519 (1959).


\bibitem{Leblond} J. M. L\'evy-Leblond, 
{\it Une nouvelle limite non-relativiste du group de Poincar\'e}, 
Ann. Inst. H. Poincar\'e {\bf 3}, 1 (1965).

\bibitem{Sen} V. D. Sen Gupta, 
{\it On an analogue of the Galileo group}, 
Il Nuovo Cimento {\bf 54}, 512 (1966).

\bibitem{Bacry} H. Bacry and J. M L\'evy-Leblond, 
{\it Possible kinematics},
J. Math. Phys. {\bf 9}, 1605 (1967).

\bibitem{Nuyts} H. Bacry and J. Nuyts,
{\it Classification of ten-dimensional kinematical groups with space
isotropy},
J. Math. Phys. {\bf 27}, 2455 (1986).

\bibitem{Minguzzi1} E. Minguzzi,
{\it Classical aspects of lightlike dimensional reduction},
Class.Quant. Grav. {\bf 23}, 7085 (2006), gr-qc/0610011.

\bibitem{Minguzzi2} E. Minguzzi,
{\it Eisenhart's theorem and the causal simplicity of Eisenhart's spacetime},
Class. Quant. Grav. {\bf 24}, 2781 (2007), gr-qc/0612014.

\bibitem{CohenGlashow} A.G. Cohen and S.L. Glashow,
{\it Very special relativity}, 
Phys. Rev. Lett.  {\bf 97} (2006) 021601, hep-ph/0601236.

\bibitem{GibbonsGomisPope} G.W. Gibbons, J. Gomis and C.N. Pope,
{\it General very special relativity is Finsler geometry}, 
Phys. Rev. {\bf D76}, 081701 (2007), 
arXiv:0707.2174 [hep-th].














\end{thebibliography}
\end{document}